\documentclass[twocolumn,aps,prl]{revtex4-2}
\usepackage{times}
\usepackage{amsmath,amsfonts,amssymb}
\usepackage{mathrsfs}
\usepackage{soul,bm,array,graphicx,bbold,multirow}
\usepackage[normalem]{ulem}
\usepackage[makeroom]{cancel}
\usepackage[usenames,dvipsnames]{xcolor}
\usepackage[colorlinks=true,citecolor=blue,urlcolor=blue,linkcolor=blue]{hyperref}
\usepackage{pdfpages}
\usepackage{lipsum}
\usepackage{multicol-2019-10-01}
 \usepackage[no-test-for-array]{nicematrix}
\usepackage{pifont}
\makeatletter
\AtBeginDocument{\let\LS@rot\@undefined}
\makeatother
\newcolumntype{x}[1]{>{\centering\let\newline\\\arraybackslash\hspace{0pt}}p{#1}}

\DeclareMathAlphabet{\mathbbold}{U}{bbold}{m}{n}

\def\bra#1{\left<{#1}\right|}				
\def\ket#1{\left|{#1}\right>}				

\newcounter{subeqn} %

\makeatletter
\@addtoreset{subeqn}{equation}
\makeatother

\begin{document}
\pdfoptionpdfminorversion=5
	
\title{Anomalous Scaling Behaviors of the Green’s Function in Critical Skin Effects}
	
\author{Yifei Yi$^1$}
\email[Corresponding author: ]{2946304594@qq.com}
\author{Zhesen Yang$^{2,3}$}
\email[Corresponding author: ]{yangzs@xmu.edu.cn}

\affiliation{$^1$College of Physics and Electronic Engineering, Sichuan Normal University, Chengdu, Sichuan 610066, China}
\affiliation{$^2$Department of Physics, Xiamen University, Xiamen 361005, Fujian Province, China}
\affiliation{$^3$Asia Paciﬁc Center for Theoretical Physics, Pohang 37673, Korea}

\date{\today}
	
\begin{abstract}
We study the Green’s functions in non-Hermitian systems exhibiting the critical non-Hermitian skin effect (critical NHSE) using a double-chain Hatano-Nelson model with inter-chain coupling $\Delta$. 
For $\Delta=0$, the system decouples into two independent chains, and the Green’s functions follow predictable patterns based on the GBZ theory. 
For small $\Delta$ ($\Delta=1/10000$), in conventional regions with trivial OBC spectral winding numbers, inter-chain coupling induces a zigzag scaling structure in Green’s functions due to competition between the two chains, explainable by first-order perturbation theory. 
In anomalous regions with non-trivial winding numbers, Green’s functions match GBZ predictions in the bulk but diverge near boundaries, with residual contributions from the $\Delta=0$ GBZ accounting for the deviations. 
These results reveal the unique non-perturbative features of critical NHSE and highlight the limitations of GBZ theory in capturing finite-size and boundary effects, emphasizing the need to consider both bulk and boundary dynamics in such systems.
\end{abstract}
	
\maketitle

Non-Hermitian systems and the non-Hermitian skin effect (NHSE) have garnered substantial attention in recent years, driven by their rich and unconventional physical phenomena~\cite{ashida2020non,yao2018edge,song2019non,okuma2020topological,
zhang2020correspondence,mcdonald2020exponentially,yi2020non,longhi2022self,
longhi2022non,brandenbourger2019non,xiao2020non,helbig2020generalized,
hofmann2020reciprocal,ghatak2020observation,zhang2021observation,chen2021realization,
zhang2021acoustic,liang2022dynamic,yao2018edge,yao2018non,kunst2018biorthogonal,
lee2019anatomy,martinez2018non,ghatak2020observation,xiao2020non,helbig2020generalized,
liang2022dynamic,ding2022non,leykam2017edge,shen2018topological,kawabata2018anomalous,
luo2019higher,ezawa2019electric,xue2020non,ao2020topological,yao2018edge,yokomizo2019non,lee2019hybrid,kawabata2020higher,okugawa2020second,fu2021non,
zhang2022universal,li2022gain,zhu2022hybrid,wang2022non,lee2019hybrid,
zou2021observation,zhang2021observation,kawabata2021topological}. 
Among the diverse manifestations of NHSE, the critical non-Hermitian skin effect (critical NHSE) stands out for its unique characteristics~\cite{li2020critical,yokomizo2021scaling,brandenbourger2019non,
schomerus2020nonreciprocal,PhysRevB.111.144307}: in the thermodynamic limit, the open-boundary condition (OBC) spectrum and eigenstates undergo discontinuous transitions across a critical point~\cite{yang2020non}; and in finite-sized systems, eigenstates also exhibit anomalous scaling behaviors~\cite{li2020critical,yokomizo2021scaling}. 

Green’s functions, which quantify a system’s response to external perturbations, are foundational to understanding dynamical processes. 
For non-Hermitian systems, formulas for OBC Green’s functions based on the generalized Brillouin zone (GBZ) have been developed for both single-band and multi-band systems~\cite{PhysRevB.103.L241408,PhysRevB.105.045122,Wanjura2020Topological,PhysRevLett.126.176601,PhysRevB.107.115412,PhysRevResearch.5.043073}. 
However, for finite-sized critical NHSE, it is natural to ask: Are there features of Green’s functions that lie beyond the predictive power of GBZ theory? 
This forms the first motivation for our work.

\begin{figure}[t]
	\centerline{\includegraphics[height=6.6cm]{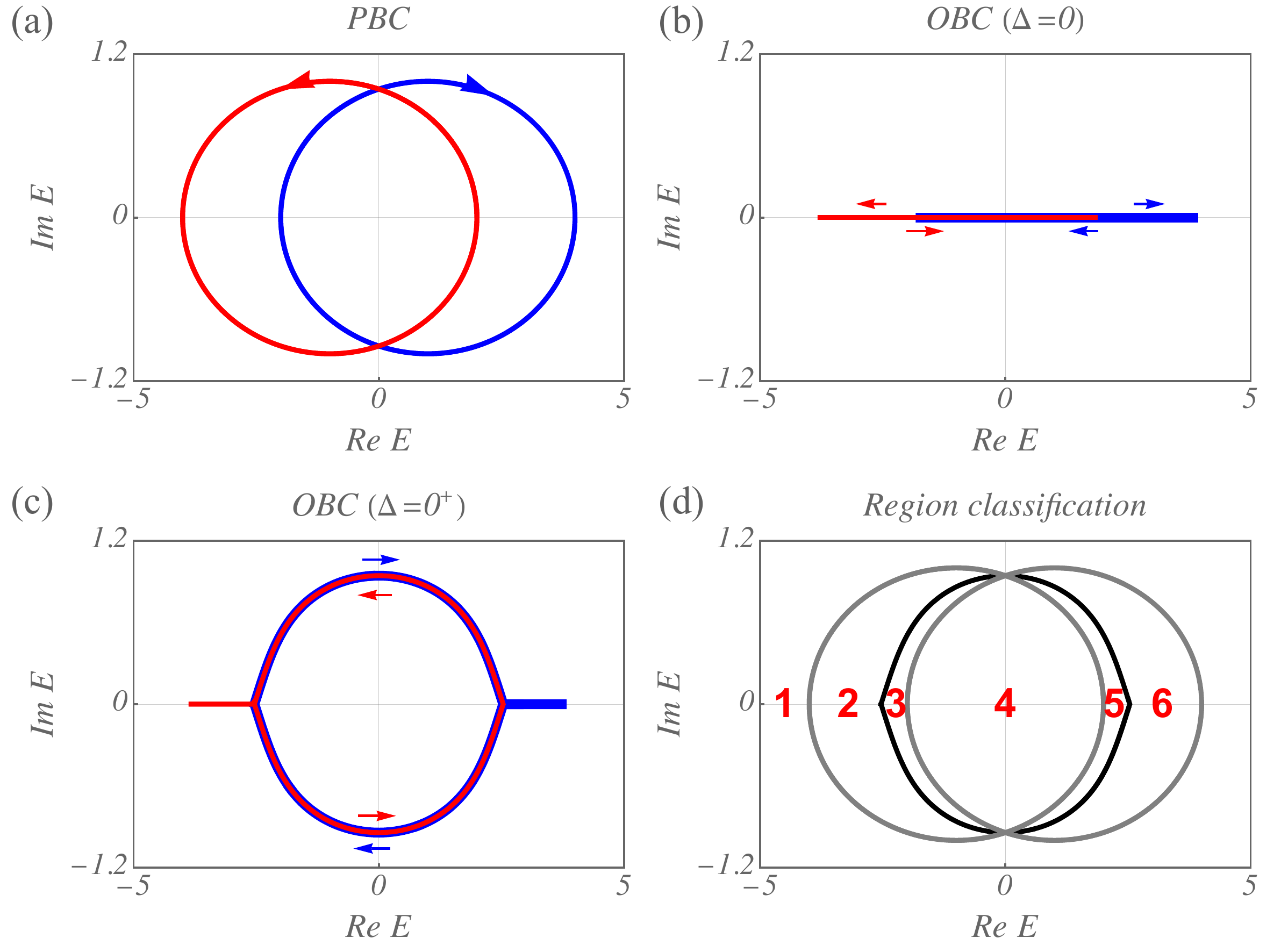}}
\caption{(a)-(c) Energy spectra of sublattices A (blue) and B (red) under periodic boundary conditions (PBC) and open boundary conditions (OBC) for $\Delta=0$ and $\Delta=0^+$, mapped from the Brillouin zone (BZ) and generalized Brillouin zones (GBZs). Arrows indicate winding directions of contour-clock mappings. (d) Classification of the complex frequency plane into six regions based on spectral winding numbers. 
Parameters are set to be $t_0=1, t_{-1}=2, t_1=1, w_0=-1, w_{-1}=1, w_1=2$.
\label{fig2}}
\end{figure}

A second motivation stems from recent experimental and theoretical advancements enabling the probing of complex-frequency Green’s functions across the entire complex plane~\cite{2024arXiv241112577H,2025arXiv250310359H}. 
For the critical NHSE, for example, as illustrated in Fig.~\ref{fig2}, certain regions (e.g., regions 3, 4, and 5) display non-zero OBC spectral winding numbers for individual bands, even when the total winding number vanishes. 
We aim to determine whether such non-trivial regions leave distinctive signatures in the Green’s functions.

Finally, critical NHSE is inherently non-perturbative, prompting us to explore: How do these non-perturbative behaviors manifest in Green’s functions? 
Can perturbation theory still be applied to describe these functions, at least in specific regions?

In this work, we investigate Green’s functions in non-Hermitian systems exhibiting critical NHSE, with a focus on their perturbative versus non-perturbative properties as well as scaling behaviors. 
Using a double-chain Hatano-Nelson model with inter-chain coupling $\Delta$, we analyze the Green’s function \(G(\omega) = (\omega - H)^{-1}\) as the driving frequency $\omega$ varies across the complex plane, comparing theoretical predictions from GBZ theory with numerical simulations.

Our key findings reveal contrasting behaviors in two types of regions within the complex plane (Fig. 1(d)): In conventional regions (with trivial OBC spectral winding numbers, i.e., regions 1, 2, and 6), inter-chain coupling induces a zigzag scaling structure in Green’s functions, arising from competition between contributions of the two decoupled chains—an effect explainable by first-order perturbation theory. 
In anomalous regions (with non-zero winding numbers, i.e., regions 3, 4, and 5), Green’s functions align with GBZ predictions in the bulk but diverge near boundaries, with residual contributions from the $\Delta=0$ GBZ accounting for these deviations. 
These results highlight the unique non-perturbative features of critical NHSE and underscore the limitations of GBZ theory in capturing finite-size and boundary effects, emphasizing the need to incorporate both bulk and boundary dynamics in such systems.

The Bloch Hamiltonian we considered in this work is given by:
\begin{equation}
\begin{aligned}
H(\beta)&=\left(\begin{array}{cc}
H_A(\beta) & \Delta \\
\Delta & H_B(\beta)
\end{array}\right),\\
&=\left(\begin{array}{cc}
t_0+t_{-1} / \beta+t_1 \beta & \Delta \\
\Delta & w_0+w_{-1} / \beta+w_1 \beta
\end{array}\right).
\label{E1}
\end{aligned}
\end{equation}
where $\beta\equiv e^{ik}$ (with $k$ denoting the wave vector), and 
$A,B$ label the two sublattices. The parameters used in our simulations are fixed as \( t_0 = 1 \), \( t_{-1} = 2 \), \( t_1 = 1 \), \( w_0 = -1 \), \( w_{-1} = 1 \), \( w_1 = 2 \).
As established in previous studies, this model exhibits the critical NHSE~\cite{li2020critical,yang2020non,PhysRevResearch.5.043073}. 
A key signature of this behavior is the discontinuous transition of the GBZ from $\Delta=0$ to $\Delta=0^+$ in the thermodynamic limit ($N\rightarrow\infty$). 
Since the Hamiltonian retains the same eigenvalues $E_A(\beta)$ and $E_B(\beta)$ for both $\Delta=0$ and $\Delta=0^+$, the discontinuity in the GBZ induces a discontinuous OBC spectrum, as illustrated in Fig.~\ref{fig2}.
For instance, mapping the Brillouin zone (BZ, labeled $C_1$), the GBZ for $E_A$ at $\Delta=0$ (labeled $C_A^0$), and the GBZ for $E_A$ at $\Delta=0^+$ (labeled $C_A^{0^+}$) onto $E_A(\beta)$ yields the blue spectra shown in Figs.~\ref{fig2} (a), (b), and (c), respectively. 
The arrows in each spectrum indicate the winding direction of the contour-clock mapping. 
For example, for the BZ, the mapping starts at $E_A(\beta=e^{i 0})$, proceeds to $E_A(\beta=e^{i \pi})$, and ends at $E_A(\beta=e^{i 2\pi})$. 
Specifically, in the OBC spectrum for $\Delta=0^+$ (Figs. ~\ref{fig2} (c)), a region with a winding number of "$-1$" emerges, distinct from the OBC with $\Delta=0$ case, where in all regions, the winding number is zero.
Similarly, mapping $C_1$, $C_B^0$ (GBZ for $E_B$ at $\Delta=0$), and $C_B^{0^+}$ (GBZ for $E_B$ at $\Delta=0^+$) onto $E_B(\beta)$ generates the red spectra in Figs.~\ref{fig2} (a)–(c).
An identical "$+1$" winding number region appears, resulting in a total zero winding number in the entire complex plane.

Based on these non-zero spectral winding numbers, the complex plane is partitioned into distinct regions as shown in Fig.~\ref{fig2} (d), with a summary of winding numbers for each region provided in Tab.~\ref{T1}. 
This work focuses on understanding the logarithmic scaling and perturbative vs non-perturbative behaviors of the OBC Green’s function across these regions for $\Delta=0$ and $\Delta=1/10000$. 
The lattice length used in simulations is $N=500$.

We first focus on the scenario where $\Delta=0$, in which the double-chain Hatano-Nelson model decouples into two independent one-dimensional chains (Chain A and Chain B). 
For each chain (treated as a single-band system), the matrix elements of the Green’s function, \( G^{\alpha\alpha}_{ij}(\omega,\Delta=0) \), describing the response at site $i$ of sublattice $\alpha$ to an external drive at site $j$ of the same sublattice with frequency $\omega$, can be well approximated by the integral formula (GBZ formula)~\cite{PhysRevB.103.L241408}:
\begin{equation}
	G_{ij}^{\alpha\alpha}(\omega,\Delta=0) \simeq \frac{1}{2\pi i}\oint_{C_{\alpha}^0} \frac{d\beta}{\beta} \frac{\beta^{i-j}}{\omega - H_{\alpha}(\beta)},\label{E4}
\end{equation}
where $C_{\alpha}^0$ denotes the GBZ for chain $\alpha$ ($\alpha=A,B$). 
From the GBZ condition, one derives  $C_A^0=\sqrt{t_{-1}/t_1}e^{i k}$ and $C_B^0=\sqrt{w_{-1}/w_1}e^{i k}$.
For any frequency $\omega$ not belonging to the OBC spectrum, the spectral winding number for each band vanishes. 
By the argument theorem, combined with $N_\alpha^{\rm poles}=1$ ($N_\alpha^{\rm poles}=1$ is the order of the pole at $\beta=0$), we find $N_\alpha^{\rm zeros}=\nu^{\rm OBC(\Delta=0)}_\alpha(\omega)+N_\alpha^{\rm poles}=1$, where $N_\alpha^{\rm zeros}$ is the number of zeros and $\nu^{\rm OBC(\Delta=0)}_\alpha(\omega)$ is the OBC spectral winding number.
This implies $C_{\alpha}^0$ encloses exactly one root of $\omega-H_\alpha(\beta)=0$ for any given $\omega$. 
Consequently, the Green’s function simplifies to: 
\begin{align}\begin{aligned}
	G^{\alpha\alpha}_{i j}(\omega,\Delta=0)\simeq& A^\alpha_0(\beta^\alpha)\left[\beta^{\alpha}(\omega)\right]^{i-j}\\ =&\begin{cases}A^\alpha_0(\beta^\alpha_1)\left[\beta^{\alpha}_1(\omega)\right]^{i-j}, & i > j ; \\[1em]A^\alpha_0(\beta^\alpha_2)\left[\beta^{\alpha}_{2}(\omega)\right]^{i-j}, & i < j .\end{cases}\label{singleband2}
\end{aligned}\end{align}
Here $A^\alpha_0(\beta^\alpha_m)=\partial_\beta H_\alpha(\beta)|_{\beta=\beta^\alpha_m}$, and $\beta_m^\alpha(\omega)$ are the roots of $\omega-H_\alpha(\beta)=0$, ordered by their magnitudes such that $|\beta_1^\alpha(\omega)|\leq |\beta_2^\alpha(\omega)|$. 
Given these roots, the asymptotic behavior of the OBC Green’s function is readily determined: (i) If $|\beta^\alpha_1(\omega)|<1$ ($|\beta^\alpha_1(\omega)|>1$), rightward attenuation (amplification) occurs; (ii) If $|\beta^\alpha_2(\omega)|<1$ ($|\beta^\alpha_2(\omega)|>1$), leftward amplification (attenuation) occurs.

\begin{figure}[t]
	\centerline{\includegraphics[width=1\columnwidth]{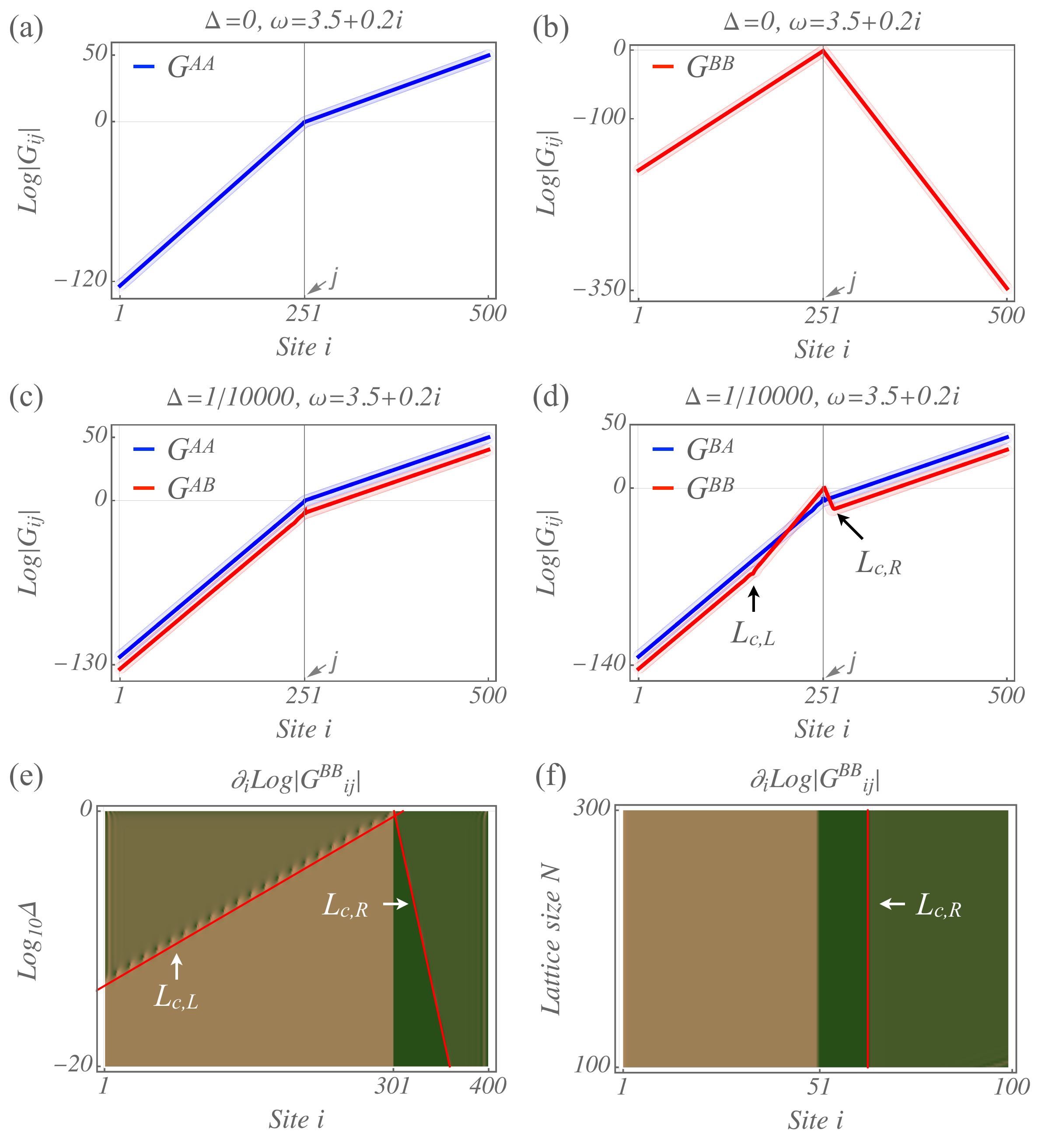}}
	\caption{Green’s functions calculated numerically (light color) and from the GBZ integral formula (solid lines) for different inter-chain couplings. (a)-(b) For $\Delta=0$ and $\omega=3.5+0.2i$ (in region 6), the system decouples into two independent chains, with Green’s functions showing distinct asymptotic behaviors (e.g., rightward amplification for $G^{AA}$). (c)-(d) For $\Delta=1/10000$ and the same $\omega$, inter-chain coupling induces a zigzag scaling structure in $G^{BB}$, with transition points $L_{c,L}$ (left) and $L_{c,R}$ (right) marked. (e)-(f) Dependence of $L_c$ on coupling strength $\Delta$ and lattice size $N$, confirming theoretical predictions. 
		\label{fig3}}
\end{figure}

\newcolumntype{C}[1]{>{\centering\let\newline\\\arraybackslash\hspace{0pt}}m{#1}}
\renewcommand\arraystretch{1.5}
\begin{table}[t]
	\caption{\label{T1}Summary of the spectral winding number in different regions shown in Fig.~\ref{fig2} (d). Here $\nu^{\rm \Delta=0}_\alpha$ and $\nu^{\rm \Delta=0}_\alpha$ represent $\nu^{\rm OBC(\Delta=0)}_\alpha$ and $\nu^{\rm OBC(\Delta=0)}_\alpha$, respectively.}
	\label{t1}
	\begin{tabular*}{7.233cm}{|C{0.3cm}|C{2cm}|C{2cm}|C{2.3cm}|}
		\hline
		\multirow{2}{*}& \multicolumn{3}{c|}{Spectral winding number}   \\   \hline          
		& $(\nu^{\rm PBC}_A,\nu^{\rm PBC}_B)$  & $(\nu^{\rm \Delta=0}_A,\nu^{\rm \Delta=0}_B)$ & $(\nu^{\rm \Delta=0^+}_A,\nu^{\rm \Delta=0^+}_B)$        \\ \hline   
		1  &$(0,0)$   &$(0,0)$ & $(0,0)$          \\ \hline   
		2  &$(0,1)$   &$(0,0)$ & $(0,0)$       \\ \hline   
		3 &$(0,1)$   &$(0,0)$ & $(-1,1)$      \\ \hline   
		4 &$(-1,1)$   &$(0,0)$ & $(-1,1)$       \\ \hline       
		5  &$(-1,0)$   &$(0,0)$ & $(-1,1)$       \\ \hline   
		6  &$(-1,0)$   &$(0,0)$ & $(0,0)$    \\ \hline   
	\end{tabular*}
\end{table}
As shown in Fig.~\ref{fig3} (a) and (b), results from Eq.~\ref{E4} (solid lines) agree well with direct numerical calculations of $G(\omega)=(\omega-H)^{-1}$ (light color circles), where $\omega$ is chosen from region 6 being equal to $3.5+0.2i$. 
Notably, only the rightward direction of $G^{AA}$ exhibits amplification. 
This behavior is determined by the PBC spectral winding number of $\omega$. 
Taking sublattice $A$ as an example: from Tab.~\ref{T1}, region 6 has $\nu^{\rm PBC}_A(\omega)=-1$. 
By the argument theorem with $N_A^{\rm poles}=1$ and $\nu^{\rm PBC}_A(\omega)=-1$, we find $N_A^{\rm zeros}=0$. 
This implies no roots of $\omega-H_A(\beta)=0$ lie within the BZ, so all roots have magnitudes greater than 1.
As $|\beta_1^A(\omega)|>1$ and  $|\beta_2^A(\omega)|>1$, the rightward amplification and leftward attenuation observed in Fig.~\ref{fig2} (a).
Similar analyses apply to other sublattices and regions. Finally, we note that for $\Delta=0$, no distinction exists between regions 2 and 3, or between regions 5 and 6.

We now turn to the scenario with a small inter-chain coupling, $\Delta=1/10000$. 
As summarized in Tab.~\ref{T1}, the complex frequency plane can be partitioned into conventional regions (regions 1, 2, and 6) with trivial OBC spectral winding numbers, and anomalous regions (regions 3, 4, and 5) with non-trivial OBC spectral winding numbers.

We first analyze the conventional regions, using the same frequency point as in  Fig.~\ref{fig3} (a) for illustration. 
Numerical calculations reveal two key features when inter-chain coupling is introduced: 
(i) Universal Asymptotic Behavior: for sufficiently large distances from the excitation, all components of the OBC Green’s function converge to the same scaling behavior. 
(ii) Zigzag Scaling Structure: 
closer to the excitation, the Green’s function, e.g., $G^{BB}$ in Fig.~\ref{fig3}(c), exhibits a zigzag pattern with two additional logarithmic scaling transition points: $L_{c,L}$ (left) and $L_{c,R}$ (right) (beyond the immediate excitation site). 
To characterize these transitions, we calculate $\partial_i\log |G^{BB}_{ij}|$ as a function of lattice size 
$N$ and coupling strength $\Delta$. 
As shown in Figs.~\ref{fig3} (e)–(f), $L_c$ (the position of the transition) increases as $\Delta$ decreases but remains invariant with increasing $N$.

To understand these features, we start from the GBZ formula for multi-band Green’s functions~\cite{PhysRevResearch.5.043073}:
\begin{equation}
	G_{ij}^{\alpha\gamma}(\omega)\simeq \frac{1}{2\pi i}\sum_{n=1}^m \oint_{C_n} \frac{d \beta}{\beta} \frac{\left\langle \alpha | R_n(\beta)\right\rangle\left\langle L_n(\beta) | \gamma\right\rangle}{\omega-E_n(\beta)}\cdot\beta^{i-j} ,\label{multiband}
\end{equation}
where $\alpha,\gamma$ represent the sublattice/spin/orbital degrees of freedom. 
$\ket{R_n( \beta)}$ and $\bra{L_n(\beta)}$ are biorthogonal eigenstates of $H(\beta)$ with energy $E_n(\beta)$, satisfying $H(\beta)\ket{R_n(\beta)}=E_n(\beta)\ket{R_n(\beta)}$ and $\bra{L_n (\beta)}H(\beta)=E_n(\beta)\bra{L_n(\beta)}$, respectively.
$C_n$ is the sub-GBZ associated with $E_n(\beta)$. 
For $\Delta=0^+$, this simplifies: 
\begin{equation}
	G_{ij}^{\alpha\gamma}(\omega,\Delta=0^+)\simeq \frac{1}{2\pi i} \oint_{C_\alpha^{0^+}} \frac{d \beta}{\beta} \frac{\delta_{\alpha\gamma} \beta^{i-j}}{\omega-H_\alpha(\beta)} ,\label{multiband1}
\end{equation}
where $C_{\alpha}^{0^+}$ is the sub-GBZ for sublattice $\alpha$ at $\Delta=0^+$. 
For conventional regions ($\nu_\alpha^{\rm OBC(\Delta=0^+)}=0$), the argument theorem ensures that $C_{\alpha}^0$ and $C_{\alpha}^{0^+}$ enclose the same root of $\omega-H_\alpha(\beta)=0$. 
Additionally, the GBZ condition for $\Delta=0^+$ (i.e., $|\beta_2|=|\beta_3|$, where $\beta_2$, $\beta_3$ are the middle two roots of the characteristic equation ${\rm{det}}\left[E-H(\beta)\right]=0$) confirms that both contours enclose the smaller root $\beta^\alpha_1$, defined in Eq.~\ref{singleband2}. 
Thus 
\begin{equation}
	G_{ij}^{\alpha\gamma}(\omega,\Delta=0)=G_{ij}^{\alpha\gamma}(\omega,\Delta=0^+),~\omega\in {\rm region~1,2,6}
\end{equation}
From this argument, one can realize that Green's function in the $\Delta=0^+$ limit cannot explain the zigzag scaling behaviors observed in Fig.~\ref{fig3}. 

For small $\Delta$, the eigenvalues and eigenstates of $H(\beta)$ admit a perturbative expansion. 
To linear order:
\begin{equation}\begin{aligned}
	E_A^{(1)}(\beta)=H_A(\beta),~~~&|R^{(1)}_A(\beta)\rangle=\left(1,\Delta/H_{AB}(\beta)\right)^t\\
	&\langle L_A^{(1)}|=\left(1,\Delta/H_{AB}(\beta)\right)\\
	E_B^{(1)}(\beta)=H_B(\beta),~~~&|R^{(1)}_B(\beta)\rangle=\left(-\Delta/H_{AB}(\beta),1\right)^t\\
	&\langle L_B^{(1)}|=\left(-\Delta/H_{AB}(\beta),1\right).
\end{aligned}\end{equation}
Here $H_{AB}(\beta)=H_A(\beta)-H_B(\beta)$. 
Substituting these into Eq.~\ref{multiband}, yields the Green’s function approximations: 
\begin{equation}\begin{aligned}
		&G_{ij}^{AA}(\omega,\Delta)\simeq G_{ij}^{AA}(\omega,0^+)+\frac{\Delta^2}{\left[H_{AB}(\beta^B)\right]^2}G_{ij}^{BB}(\omega,0^+),\\
		&G_{ij}^{BB}(\omega,\Delta)\simeq G_{ij}^{BB}(\omega,0^+)+\frac{\Delta^2}{\left[H_{AB}(\beta^A)\right]^2}G_{ij}^{AA}(\omega,0^+).\label{E8}
\end{aligned}\end{equation}
And 
\begin{equation}\begin{aligned}
	&G^{AB}_{ij}(\omega,\Delta)=G^{BA}_{ij}(\omega,\Delta)\\
	=&\frac{\Delta}{H_{AB}(\beta^A)}G_{ij}^{AA}(\omega,0^+)-\frac{\Delta}{H_{AB}(\beta^B)}G_{ij}^{BB}(\omega,0^+). 
\end{aligned}\end{equation}
Here $H_{AB}(\beta^\alpha)=H_{AB}(\beta^\alpha_1)$ for $i>j$, and $H_{AB}(\beta^\alpha_2)$ for $i<j$.
This perturbation theory fully explains the observed scaling features:
The zigzag structure arises from competition between $G_{ij}^{AA}(\omega,0^+)$ and $G_{ij}^{BB}(\omega,0^+)$.  
As seen in Figs.~\ref{fig3} (a)–(b), $G^{AA}$ decays more slowly or amplifies more rapidly than $G^{BB}$ in both directions. 
For short distances, $G^{BB}$ dominates due to its larger coefficient, but $G^{AA}$ takes over at large distances.
The transition points $L_c$ satisfy $|G_{ij}^{BB}(\omega,0^+)|\sim |\Delta^2/\left[H_{AB}(\beta^A)\right]^2G_{ij}^{AA}(\omega,0^+)|$, leading to 
\begin{equation}
 L_c=i-j=\frac{2\log\Delta+|A_0^B(\beta^B)[H_{AB}(\beta^B)]^{-2}|/|A_0^A(\beta^A)|}{\log[|\beta^A|/|\beta^B|]},
\end{equation}
where $A_0^\alpha(\beta^\alpha)$ is defined in Eq.~\ref{singleband2}.  
This formula agrees well with numerical results [Fig.~\ref{fig3} (e)–(f)].

\begin{figure}[t]
	\centerline{\includegraphics[height=8cm]{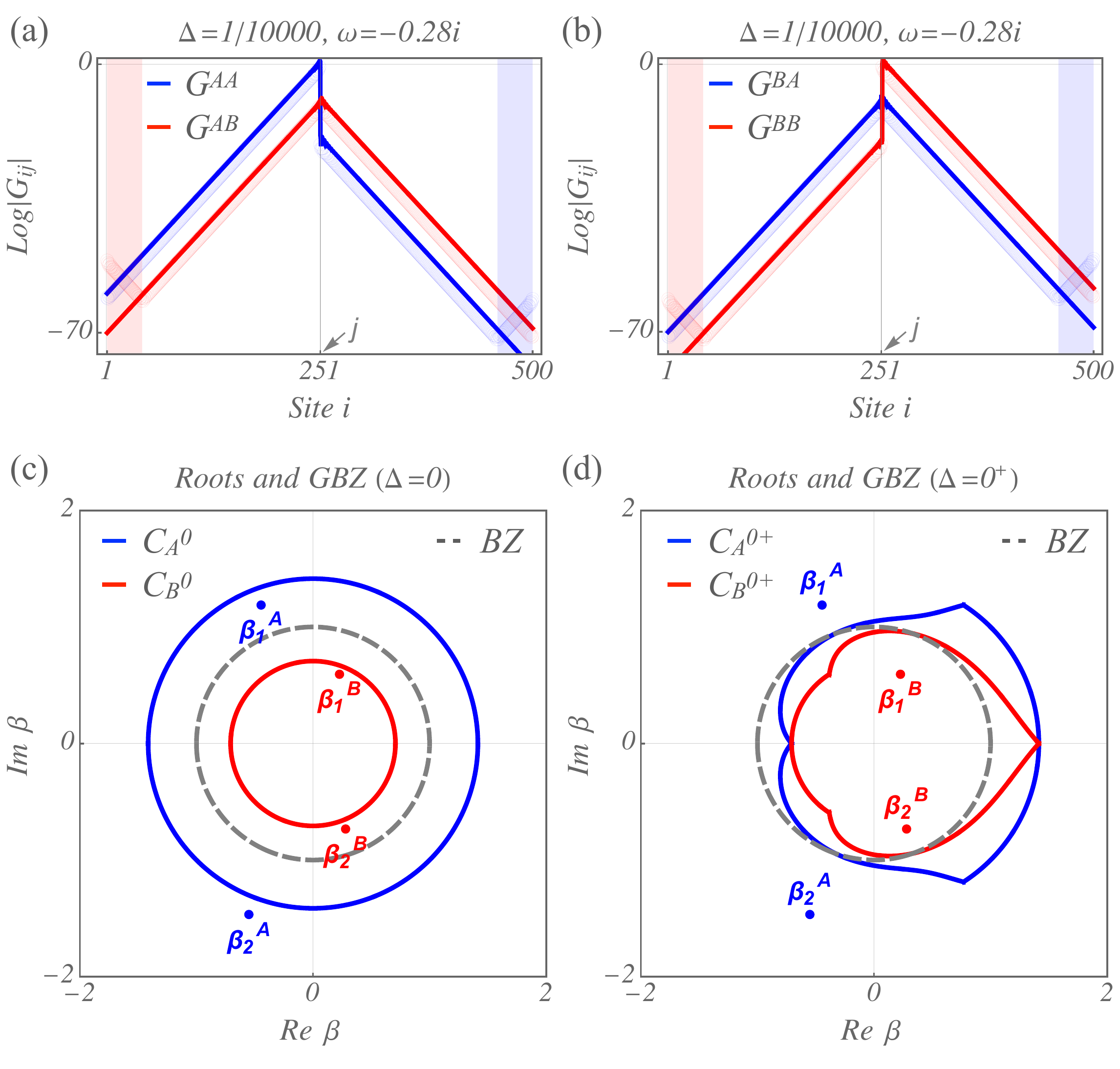}}
	\caption{Green’s functions and root distributions for $\omega= -0.28i$ (in region 4). (a)-(b) Numerical results for $\Delta=1/10000$ show distinct behaviors with the GBZ prediction near the boundary. (c)-(d) Root positions and generalized Brillouin zones (GBZs) for $\Delta=0$ and $\Delta=0^+$, where $C_\alpha^0$ and $C_\alpha^{0^+}$ enclose different roots, leading to boundary deviations in Green’s functions from GBZ predictions.
		\label{fig4}}
\end{figure}

We now focus on the anomalous regions with $\Delta=1/10000$, where the driving frequency $\omega$
lies within regions with non-zero OBC spectral winding numbers (i.e., regions 3–5). 
For concreteness, we set $\omega=-0.28 i$ in our discussion. 
As shown in Figs.~{\ref{fig4}} (a) and (b), the zigzag scaling structure of the Green’s function persists in these regions. 
However, a critical discrepancy emerges: while the Green’s function agrees well with GBZ formula predictions in the bulk, significant deviations arise near the boundaries. 
This raises two key questions: 
What causes this boundary divergence? 
Why does GBZ theory fail to capture the scaling behavior near boundaries in anomalous regions?

To address these, we first analyze the $\Delta=0^+$ limit. From Tab.~\ref{T1}, the OBC spectral winding numbers satisfy $\nu_A^{\rm OBC(\Delta=0^+)}=-1$ and $\nu_B^{\rm OBC(\Delta=0^+)}=+1$. 
By the argument theorem, this implies:
(i) The sub-GBZ $C_A^{0^+}$ (for sublattice $A$ at $\Delta=0^+$) encloses zero roots of $\omega-H_A(\beta)=0$; 
(ii) The sub-GBZ $C_B^{0^+}$ (for sublattice $B$ at $\Delta=0^+$) encloses two roots of $\omega-H_B(\beta)=0$. 
In contrast, for $\Delta=0$, both $C_A^{0}$ and $C_B^{0}$ enclose exactly one root each (as established in the previous discussion). 
This fundamental difference leads to a discontinuous jump in the Green’s function at $\Delta=0$ in the thermodynamic limit:
\begin{equation}
	G_{ij}^{\alpha\gamma}(\omega,\Delta=0)\neq G_{ij}^{\alpha\gamma}(\omega,\Delta=0^+).~\omega\in {\rm region~3,4,5}
\end{equation}
To further clarify the scaling behavior of $G_{ij}^{\alpha\gamma}(\omega,\Delta=0^+)$ for $\omega=-0.28i$, we plot the sub-GBZs and corresponding roots in Fig.~\ref{fig4} (d). 
Here, $C_A^{0^+}$ encloses no roots, and $C_B^{0^+}$ encloses two roots as we expected. 
By the residue theorem, this implies $G_{i>j,j}^{AA}(\omega,\Delta=0^+)=G_{i<j,j}^{BB}(\omega,\Delta=0^+)=0$ exactly. 
Additionally, all roots inside $C_B^{0^+}$ (or outside $C_A^{0^+}$) have magnitudes less than 1 (or greater than 1), causing $G^{AA}$ and $G^{BB}$ to decay along the left and right directions, respectively. 
However, similar to the conventional regions, the $\Delta=0^+$ limit fails to explain the observed zigzag behavior.
First-order perturbation theory (Eq.~\ref{E8}) offers partial insight: inter-chain coupling introduces small non-zero contributions proportional to $\Delta^2$ on the right side of $G_{ij}^{AA}(\omega,\Delta)$
and the left side of $G_{ij}^{BB}(\omega,\Delta)$, explaining the jump in the Green’s function at the excitation site. 
Notably, though, both $G_{ij}^{AA}(\omega,\Delta)$ and $G_{ij}^{BB}(\omega,\Delta)$ decay along their respective directions, leaving the observed amplification in the final results unexplained. 
This limitation extends to the original GBZ formula Eq.~\ref{multiband}, as the fixed relationship between sub-GBZs and pole positions prohibits amplification behaviors.

The key to resolving this anomaly lies in residual contributions from the $\Delta=0$ GBZ. 
Taking $G^{BB}$ as an example: near the left boundary, the scaling behavior is dominated by $\beta_2^B$, a root being outside of $C_B^{0}$ (GBZ for $\Delta=0$) but inside of $C_B^{0^+}$ (GBZ for $\Delta=0^+$) [Fig.~\ref{fig4}(c)]. 
This indicates that the zigzag transitions in anomalous regions arise from competition between roots $\beta_2^B$ and $\beta_{1/2}^A$.
Similar analyses apply to other anomalous regions. 
In regions 3 and 5, while boundary amplification is absent, the Green’s function still deviates from GBZ predictions, underscoring the critical role of residual contributions from the $\Delta=0$ GBZ in capturing finite-size boundary effects unique to critical NHSE.

A key question emerges from our analysis: why does the GBZ formula fail to capture the Green’s function behavior in anomalous regions? 
The root cause lies in the foundational assumption of GBZ theory, which is derived from bulk properties and inherently neglects boundary effects—even those of perturbative significance.
This parallels the Hermitian case, where GBZ theory yields accurate bulk predictions but deviates near boundaries when the system is excited close to its edges.
In anomalous regions, the breakdown of GBZ predictions stems from residual contributions from the $\Delta=0$ GBZ, which persist despite the small inter-chain coupling ($\Delta=1/10000$). 
These contributions, though very weak, introduce a finite boundary length where the Green’s function scaling deviates from GBZ-based expectations. 
Specifically, the competition between roots enclosed by the $\Delta$ GBZ ($C^0$) and the $\Delta=0^+$ GBZ ($C^{0^+}$) drives the observed zigzag transitions near boundaries. 

In summary, we investigate the Green’s functions of non-Hermitian systems exhibiting the critical NHSE, focusing on their scaling behaviors and deviations from GBZ theory. 
Using a double-chain Hatano-Nelson model with inter-chain coupling ($\Delta$), we analyze the Green’s function \( G(\omega) = (\omega - H)^{-1} \) across the complex frequency plane. 
For $\Delta=0$, the system decouples into two independent chains, with Green’s functions following analytical predictions from GBZ integrals, their asymptotic behaviors determined by root magnitudes of \( \omega - H_\alpha(\beta) = 0 \). 
For $\Delta=1/10000$, in conventional regions (trivial OBC spectral winding, i.e., regions 1, 2, 6), inter-chain coupling induces a zigzag scaling structure in Green’s functions due to competition between contributions of the two decoupled chains, explained by first-order perturbation theory. 
In anomalous regions (non-zero winding numbers, i.e., regions 3, 4, 5), Green’s functions agree with GBZ predictions in the bulk but diverge near boundaries, with residual contributions from the $\Delta=0$ GBZ accounting for the deviations. 
These results highlight the unique non-perturbative features of critical NHSE and the limitations of GBZ theory in capturing finite-size and boundary effects.

Y.Yi was supported by the National Natural Science Foundation of China (Grant No.12304201).
Z.Yang acknowledges the support from the National Key R\&D Program of China (Grant No. 2023YFA1407500) and the National Natural Science Foundation of China (Grants No. 12322405).

\bibliography{refs1}
\bibliographystyle{apsrev4-1}
	
\end{document}